\documentclass[11pt]{article}
\usepackage{hyperref}
\pdfoutput=1

\begin{document}
\title{Trapping, focusing and sorting of microparticles through bubble streaming}
\author{Cheng Wang, Shreyas Jalikop and Sascha Hilgenfeldt \\
\\\vspace{6pt} Department of Mechanical Science and
Engineering \\ University of Illinois at Urbana-Champaign \\
1206 W Green St, Urbana, IL 61801, USA}
\maketitle
\begin{abstract}
Ultrasound-driven microbubbles rectify oscillatory motion into
strong steady streaming flow [1, 2]. In this fluid dynamics video,
we show size-dependent trapping, focusing and sorting of
microparticles, utilizing the interaction between bubble streaming
and the induced streaming around the mobile particles. This
mechanism is purely passive and is  selective for sizes and
size differentials far smaller than any scale imposed by the
device geometry.
\end{abstract}
\section*{Movie description}

The video is available in
\href{/anc/APS_video_microbubble_final_high.mpg} {High
resolution} and \href{/anc/APS_video_microbubble_final_low.mpg}
{Low resolution}. It consists of 5 movie segments. All movies are played at 30 frames per second.
The detailed descriptions of the movies are as follows:

\begin{enumerate}
\item This movie shows the motion a 5$\mu m$ tracing particle
around an oscillating microbubble at both fast and slow time
scales. The fast time scale illustrated the fast, low-amplitude
oscillatory flow; at the slow time scale, the secondary steady
streaming flow with closed streamlines is visible. The ultrasound
driving frequency is 25.5 kHz and the amplitude is 10\,V$_{rms}$.
Movie 1(a) was captured at 95238 frames per second; movie 1(b) was
captured at 30 frames per second.

\item This movie shows the size dependent characteristic
trajectories of suspended particles in bubble streaming flow. The 5$\mu m$ and
10$\mu m$ density-matched particles are drawn towards the oscillating bubbles due
to an attractive bubble/particle interaction resulting from induced streaming. Because the
particles cannot penetrate the bubbles, they are forced onto
different trajectories. The movie was captured at 250 frames per
second.

\item This movie shows that superimposing unidirectional
Poiseuille flow destroys the symmetry of the streaming flow. The differential strength of
the bubble/particle interaction now selects which particles are
transported and which are trapped near the bubbles. In 3(a),
10$\mu m$ particles are trapped upstream of the bubble,
while 2$\mu m$ and 5$\mu m$ particles are carried downstream.
Further upstream accumulation leads to the escape of the larger
particles into  narrow trajectory bundles. The movie was captured
at 200 frames per second.

\item The trapping and releasing mechanism can be further
exploited for focusing and sorting. This movie shows that
alternating microbubbles along a microchannel focuses 10$\mu m$
particles into a very narrow trajectory bundle. In 4(a), the
particles remain distributed widely as the bubbles
are not excited. In 4(b), the oscillating microbubbles successively relay the
particles through the trapping-and-releasing
mechanism. The movie was captured at 100 frames per second. In 4(c), 
we compare results from experiment and a simulation of the steady streaming flow based on Stokes flow singularities.

\item This movie shows that an oscillating bubble placed near a
T-junction sorts particles above a certain size into one specific
outlet. Optimizing ultrasound driving frequency and amplitude
allows for the selection of 5$\mu m$ particles to be trapped,
released, and sorted over 2$\mu m$ particles, which remain
unsorted. This demonstrates the flexibility and tunability of this
technique, which can overcome physical restrictions of the device
itself. Without microbubble excitation (5a), both 5$\mu m$ and
2$\mu m$ particles are equally carried to the upper and lower
outlets. In 5(b), upon exciting, the microbubble sorts the 5$\mu
m$ particles into the upper outlet. The movie was captured at 100
frames per second.

\end{enumerate}

\section*{References}
\begin{enumerate}
    \item Marmottant, P., \& Hilgenfeldt, S.  \emph{Nature}, \textbf{423}(6936), 153-156 (2003).
    \item Marmottant, P., Raven, J. P., Gardeniers, H., Bomer, J. G., \& Hilgenfeldt, S. \emph{Journal of Fluid Mechanics}, \textbf{568}, 109-118 (2006).
\end{enumerate}
\end{document}